\documentclass[a4paper,12pt]{article}
\usepackage[cp1251]{inputenc}
\usepackage[english, russian]{babel}
\usepackage[dvips]{graphicx}
\graphicspath{{.}}
\usepackage{amsfonts}
\usepackage{amsmath}

\begin{document}
\mathsurround=2pt \sloppy
%\begin{center}
\title{Effect of random anisotropy on NMR frequency shift in the polar phase of the superfluid $^3$He.}
\author { I. A. Fomin
\vspace{.5cm}\\
{\it  P. L. Kapitza Institute for Physical Problems}\\ {\it Russian
Academy of Science},
\\{\it Kosygina 2,
 119334 Moscow, Russia}}

%\end{center}
\maketitle
\begin{abstract}
The orbital anisotropy induced in the superfluid  $^3$He by the nematic aerogel is generally speaking spatially non-uniform. This anisotropy in its turn induces spatial fluctuations of the order parameter.  It is shown here that for the polar phase these fluctuations decrease the overall amplitude of the order parameter and the magnitude of the frequency shift of the transverse NMR. Different contributions to this effect are discussed and estimated. Their temperature dependencies are discussed as well.
\end{abstract}

\section{Introduction}
The polar phase of the superfluid $^3$He corresponds to the Cooper pairing with the orbital moment  $l=1$  and its projection on a chosen direction $l_z=0$.  Its order parameter can be represented as  3$\times$3 complex matrix $A_{\mu j}=\Delta_P\exp(i\varphi)d_{\mu} m_j$, where $d_{\mu}$  is a real unit vector in the spin space,  $m_j$  a unit orbital vector and  $\Delta_P$  is an overall amplitude. In the bulk liquid $^3$He the polar phase is not favorable energetically. For its possible stabilization  in the experiments \cite{AG1,Dm-pp,Parp} the \emph{nematic} aerogels were used. These are aerogels formed by straight and nearly parallel strands. Ensemble of such strands can induce the average orbital anisotropy which according to theoretical arguments  \cite{AI,SF1,fom3} has to stabilize a phase with  $l_z=0$. Two types of nematic aerogels were used - ``obninsk'' and ``nafen'' \cite{nf}. The polar phase was definitely observed in nafen \cite{Dm-pp}. Its experimental identification was partly based on the temperature dependence of the transverse NMR frequency shift $\Delta\omega$ from the Larmor frequency  $\omega_L$. In the notations of Ref.\cite{Dm-pp}  for the CV NMR $2\omega_L\Delta\omega=K\Omega^2_A$, where  $\Omega_A$ is the frequency of longitudinal resonance in the bulk A-phase. In a mean field approximation the coefficient $K$ depends on a form of the average order parameter. In particular when the d.c. magnetic field is parallel to the strands of aerogel  $K=4/3$ for the polar phase and $K=1/2$ for the A-phase, if the strong coupling corrections are neglected. .

It has to be taken into account that average distances between the strands of nematic aerogels used in the experiments \cite{AG1,Dm-pp} are of the order of the correlation length in superfluid $^3$He. As a result the anisotropy induced by these aerogels is generally non-uniform. Except for the average global anisotropy a random local anisotropy does exist. The local anisotropy induces random fluctuations of the order parameter which can effect its average  amplitude and the value of the coefficient $K$ without changing the form of the order parameter.  With respect to the experiments \cite{AG1,Dm-pp} it could mean that the polar phase was observed not only in nafen but in the ``obninsk'' aerogel as well.
\section{Random anisotropy}
In the bulk  $^3$He the temperature of transition in the superfluid state $T_c$ is the same for all three projections of the orbital momentum. Nematic aerogel lifts  the degeneracy and splits the transition. The splitting may be different in different points of the sample. In a vicinity of the $T_c$ the splitting can be included in the second order term of the expansion of the density of free energy $f_{ag}=N(0)\Lambda_{jl}(\mathbf{r})A_{\mu j}A_{\mu l}^\ast$ where $N(0)$ - the density of states and  $\Lambda_{jl}(\mathbf{r})$  is real symmetric tensor. This tensor can be represented as $\Lambda_{jl}(\mathbf{r})=\kappa_{jl}+\eta_{jl}(\mathbf{r})$, where $\kappa_{jl}=\langle\Lambda_{jl}\rangle$ is global anisotropy. The remaining part  $\eta_{jl}(\mathbf{r})$ is a local anisotropy. At a strength of its definition  $\langle\eta_{jl}\rangle=0$.  The angular brackets here and in what follows denote the ensemble averaging. We assume that aerogel is on the average axially symmetric. In a coordinate system with $z$-axis parallel to the strands tensor  $\kappa_{jl}$ is diagonal and its principal values are $\tau_{\parallel}=(1-T_{\parallel}/T)$  and   $\tau_{\perp}=(1-T_{\perp}/T)$, where $T$ is the temperature. Constants  $T_{\parallel}$ and $T_{\perp}$ are the temperatures at which change their signs eigenvalues of  $\kappa_{jl}$ corresponding to the states with  $l_z=0$ and  $l_z=\pm 1$ respectively. In a nematic aerogel $T_{\parallel}>T_{\perp}$. In these notations the gain of the free energy of $^3$He at its transition in the superfluid state can be written as:
$$
\frac{f_S-f_N}{N(0)}=[\tau_{\parallel}\hat{z}_j\hat{z}_l+\tau_{\perp}(\hat{x}_j\hat{x}_l+\hat{y}_j\hat{y}_l)+\eta_{jl}(\mathbf{r})]A_{\mu j}A_{\mu l}^\ast+\xi_s^2\left(\frac{\partial A_{\mu
l}}{\partial x_n} \frac{\partial A^*_{\mu l}}{\partial
x_n}\right)+\frac{1}{2}\sum_{s=1}^5 \beta_sI_s.                                        \eqno(1)
$$
 Here $I_s$ are invariants of the 4-th order \cite{VW}: $I_1=A_{\mu j}A_{\mu j}A_{\nu l}^*A_{\nu l}^*$, $I_2=A_{\mu j}A_{\mu j}^*A_{\nu l}A_{\nu l}^*$, $I_3=A_{\mu j}A_{\nu j}A_{\mu l}^*A_{\nu l}^*$, $I_4=A_{\mu j}A_{\nu j}^*A_{\nu l}A_{\mu l}^*$, $I_5=A_{\mu j}A_{\nu j}^*A_{\mu l}A_{\nu l}^*$, and  $\beta_1,...\beta_5$ - phenomenological coefficients. For the gradient energy for the sake of simplicity the isotropic expression is taken. Variation of  the free energy over  $A_{\mu l}^\ast$ renders an equation for the equilibrium order parameter:
$$
[\tau_{\parallel}\hat{z}_j\hat{z}_l+\tau_{\perp}(\hat{x}_j\hat{x}_l+\hat{y}_j\hat{y}_l)]A_{\mu l}-\xi_s^2\left(\frac{\partial^2 A_{\mu
j}}{\partial x_n^2}\right)+\frac{1}{2}\sum_{s=1}^5 \beta_s\frac{\partial I_s}{\partial A_{\mu j}^\ast}=-\eta_{jl}(\mathbf{r})A_{\mu l}.              \eqno(2)
$$
It will be assumed in what follows that the random anisotropy $|\eta_{jl}|$ is small in comparison with the global, which can be characterized by the relative splitting of the transition temperature $\tau_{\perp}-\tau_{\parallel}$. The random anisotropy will be treated as a perturbation in the analogy with the analysis of effect of spacial fluctuations of the transition temperature for the case of the $s$-pairing by Larkin and Ovchinnikov \cite{LO}.
The solution of the Eq.(2) has to be searched in a form $A_{\mu j}=\bar{A}_{\mu j}+a_{\mu j}$, where $\bar{A}_{\mu j}$ is the order parameter averaged over distances which are greater than the characteristic scale for variation of the random anisotropy  $\eta_{jl}(\mathbf{r})$. It is $\bar{A}_{\mu j}$, which is considered as the order parameter of the equilibrium phase for given conditions. The fluctuation  $a_{\mu j}$ vanishes at such averaging $\langle a_{\mu j}\rangle=0$. Principal corrections to the macroscopic quantities of the liquid are proportional to the averaged products of  the fluctuations  $\langle a_{\nu l} a_{\eta r}\rangle, \langle a_{\nu l}^\ast a_{\eta r}\rangle$.  Keeping in Eq. (2) terms up to the 2-nd order over $a_{\mu j}$ and $\eta_{jl}(\mathbf{r})$ and  averaging the obtained equation we arrive at:
$$
[\tau_{\parallel}\hat{z}_j\hat{z}_l+\tau_{\perp}(\hat{x}_j\hat{x}_l+\hat{y}_j\hat{y}_l)]\bar{A}_{\mu l}
+\frac{1}{2}\sum_{s=1}^5 \beta_s\frac{\partial I_s}{\partial A_{\mu j}^\ast}+
$$
$$
\frac{1}{4}\sum_{s=1}^5 \beta_s\left[\frac{\partial^3 I_s}{\partial A_{\mu j}^\ast\partial A_{\nu l}\partial A_{\eta r}} \langle a_{\nu l}a_{\eta r}\rangle+
2\frac{\partial^3 I_s}{\partial A_{\mu j}^\ast\partial A_{\nu l}^\ast\partial A_{\eta r }}\langle a_{\nu l}^\ast a_{\eta r}\rangle \right] =-\langle\eta_{jl}(\mathbf{r})a_{\mu l}\rangle-\tau_{jl}^{(1)}\bar{A}_{\mu l}.                            \eqno(3)
$$
Derivatives of the invariants  $I_s$ are taken at $A_{\mu j}=\bar{A}_{\mu j}$. In the terms, containing small factors for the order parameter $\bar{A}_{\mu j}$ a solution of the zero order equation can be used:
$$
[\tau_{\parallel}\hat{z}_j\hat{z}_l+\tau_{\perp}(\hat{x}_j\hat{x}_l+\hat{y}_j\hat{y}_l)]\bar{A}_{\mu l}
+\frac{1}{2}\sum_{s=1}^5 \beta_s\frac{\partial I_s}{\partial A_{\mu j}^\ast}=0,                                   \eqno(4)
$$
i.e. the order parameter of the polar phase $\bar{A}_{\mu j}^{0}=\Delta_0 \exp(i\varphi)d_{\mu}m_j$ with $\Delta_0^2=\frac{-\tau_{\parallel}}{\beta_{12345}}$ and $\beta_{12345}=\beta_1+\beta_2+...+\beta_5$. Collecting the first order terms we arrive at the equation for $a_{\mu j}$:
$$
[\tau_{\parallel}\hat{z}_j\hat{z}_l+\tau_{\perp}(\hat{x}_j\hat{x}_l+\hat{y}_j\hat{y}_l)]a_{\mu l}-\xi_s^2\left(\frac{\partial^2 a_{\mu
j}}{\partial x_n^2}\right)
+\frac{1}{2}\sum_{s=1}^5 \beta_s\left[\frac{\partial^2 I_s}{\partial A_{\mu j}^\ast\partial A_{\nu l}}a_{\nu l}+\frac{\partial^2 I_s}{\partial A_{\mu j}^\ast\partial A_{\nu l}^{\ast}}a_{\nu l}^{\ast}\right]=-\eta_{jl}(\mathbf{r})\bar{A}_{\mu l}.                            \eqno(5)
$$
The average order parameter here is assumed to be spatially uniform. Projections of  $a_{\mu j}$ on the spin vectors $e_{\mu},f_{\mu}$, forming together with $d_{\mu}$ orthogonal basis, satisfy homogeneous equations, which do not depend on the random anisotropy. In the region of stability of the polar phase these projections can be omitted and only projection of $a_{\mu j}$ on $d_{\mu}$ is essential. Then one can multiply Eq. (5) on $d_{\mu}$ and to solve it with respect to the orbital vector   $a_j=d_{\mu}a_{\mu j}$. Using the explicit form of the invariants $I_s$ we obtain:
$$
[\tau_{\parallel}\hat{m}_j\hat{m}_l+\tau_{\perp}(\hat{n}_j\hat{n}_l+\hat{l}_j\hat{l}_l)]a_l-\xi_s^2\left(\frac{\partial^2 a_j}{\partial x_n^2}\right)
+
$$
$$
\Delta_0^2 \left\{\beta_{13}[2m_j(a_s m_s)+a_j^\ast]+\beta_{245}[a_j+m_jm_s(a_s+a_s^{\ast})]\right\}=-\eta_{jl}(\mathbf{r})\bar{A}_l.                         \eqno(6)
$$
It is taken into account here that nematic aerogel orients the orbital vector $m_j$ parallel to its strands and two other orbital unit vectors $n_j$ and $l_j$ are introduced,  which form together with $m_j$ an orthogonal basis. Substitution $a_j=b_j+ic_j$ in the linear Eq. (6) renders separate equations for real and imaginary parts of $a_j$. The overall phase of the order parameter can be chosen so that $\bar{A}_l$ is real. Then the equation for $c_j$ is homogeneous. Its solution does not depend on the random anisotropy. The longitudinal component  $(\hat{m}_lc_l)$ is a small correction to the overall phase and  transverse component is absent within the region of stability of the polar phase. Equation for the real part has a form:
$$
-\tau_{\parallel}[\hat{m}_j(\hat{m}_lb_j)+b_l]+\tau_{\perp}[(\hat{n}_j\hat{n}_l+\hat{l}_j\hat{l}_l)]b_l-\xi_s^2\left(\frac{\partial^2 b_j}{\partial x_s^2}\right)=-\eta_{jl}\bar{A}_l.                  \eqno(7)
$$
Projections of this equation on  $\hat{m}_j, \hat{n}_j, \hat{l}_j$ renders respectively equation for the dimensionless longitudinal component  $\hat{b}_{\|}\equiv b_jm_j/\Delta_P$ :
$$
2\tau_{\|}\hat{b}+\xi_s^2\left(\frac{\partial^2 \hat{b}_{\|}}{\partial x_s^2}\right)=\eta_{zz}                  \eqno(8)
$$
and each of the two transverse $\hat{b}_{\bot 1}\equiv b_jn_j/\Delta_P$ and  $\hat{b}_{\bot 2}\equiv b_jl_j/\Delta_P$;
$$
(\tau_{\bot}-\tau_{\|})\hat{b}_{\bot\alpha}-\xi_s^2\left(\frac{\partial^2 \hat{b}_{\bot\alpha}}{\partial x_s^2}\right)=-\eta_{\alpha z},        \eqno(9)
$$
where $\alpha$ runs over two values - 1,2 or $x,y$. Linear equations (8) and (9) are solved by Fourier transformation  $\hat{b}_{\|}(\mathbf{k})=\int \hat{b}_{\|}(\mathbf{r})\exp(-i\mathbf{k}\mathbf{r})d^3r$ etc.:
$$
\hat{b}_{\|}(\mathbf{k})=-\frac{\eta_{zz}(\mathbf{k})}{\xi_s^2k^2-2\tau_{\|}}; \qquad  \hat{b}_{\bot\alpha}=-\frac{\eta_{\alpha z}(\mathbf{k})}{\xi_s^2k^2+(\tau_{\bot}-\tau_{\|})}.                                                                  \eqno(10)
$$
Principal order corrections to the value of the NMR-shift contain averages
 $\langle \hat{b}_{\|}(\mathbf{r})\hat{b}_{\|}(\mathbf{r})\rangle$ и   $\langle \hat{b}_{\bot\alpha}(\mathbf{r})\hat{b}_{\bot\alpha}(\mathbf{r})\rangle $. They do not depend on $\mathbf{r}$. The average $\langle \hat{b}_{\|}\hat{b}_{\|}\rangle$ can be expressed in terms of the correlation function  $f_{\|}(\mathbf{r})=\langle\eta_{zz}(0)\eta_{zz}(\mathbf{r})\rangle$:
$$
\langle \hat{b}_{\|}\hat{b}_{\|}\rangle=\frac{1}{8\pi\xi_s^3\sqrt{2\mid\tau_{\|}\mid}}\int d^3r f_{\|}(\mathbf{r})\exp(-r/\xi_{\|}),      \eqno(11)
$$
or in terms of its Fourier transformation  $f_{\|}(\mathbf{k})$:
$$
\langle \hat{b}_{\|}\hat{b}_{\|}\rangle=\int\frac{f_{\|}(\mathbf{k})}{(\xi_s^2k^2-2\tau_{\|})^2}\frac{d^3k}{(2\pi)^3},         \eqno(12)
$$
where $\xi_{\|}=\xi_s/\sqrt{2\mid\tau_{\|}\mid}$. On the approach to the superfluid transition temperature $T_{\|}$ this correction diverges as  $1/\sqrt{\mid\tau_{\|}\mid}$. The perturbation theory applies until this correction is small, i.e.
$$
\frac{1}{8\pi\xi_s^3\sqrt{2\mid\tau_{\|}\mid}}\int d^3r f_{\|}(\mathbf{r})\exp(-r/\xi_{\|})\ll 1.                        \eqno(13)
$$
The integral here can be estimated in a following way. Assume that correlations of $\eta_{zz}(\mathbf{r})$ decay on a characteristic  distance $R_{\|}$ and that this distance is (roughly) isotropic. At  $\mid\tau_{\|}\mid\to 0$  Ginzburg and Landau correlation length  $\xi_{\|}\to\infty$ and for sufficiently small  $\mid\tau_{\|}\mid$ a strong inequality  $R_{\|}\ll\xi_{\|}$ will be met. In this case the exponent in Eq, (13) is close to unity and the remaining integral can be estimated as  $\mid\eta_{zz}\mid^2 R_{\|}^3$. The condition of applicability of the perturbation theory  reads then as $\mid\tau_{\|}\mid\gg\left(\frac{R_{\|}^3}{\xi_s^3}\mid\eta_{zz}\mid^2\right)^2$.

In the opposite limit ($R_{\|}\gg\xi_{\|}$) convergence of the integral in Eq. (13) is secured by the exponent and the correlation function within the region of integration can be considered as a constant $f_{\|}(0)$. In this case we have the following estimation : $\langle \hat{b}_{\|}\hat{b}_{\|}\rangle\sim\frac{\mid\eta_{zz}\mid^2}{\tau_{\|}^2}$. It means that a relative contribution of longitudinal fluctuations of the order parameter in the macroscopic characteristics of the polar phase by the order of magnitude is equal to the square of the amplitude of the random part of the longitudinal component of anisotropy to the average anisotropy.

The average products of transverse components  $\hat{b}_{\bot\alpha}$ are expressed in the analogy with Eqns. (11),(12) via correlation functions
$f_{\alpha}(\mathbf{r})=\langle\eta_{\alpha z}(0)\eta_{\alpha z}(\mathbf{r})\rangle$ or their Fourier transforms $f_{\alpha}(\mathbf{k})$:
$$
\langle \hat{b}_{\bot\alpha}\hat{b}_{\bot\alpha}\rangle=\frac{1}{8\pi\xi_s^3\sqrt{\tau_{\bot}-\tau_{\|}}}\int d^3r f_{\alpha}(\mathbf{r})\exp(-r/\xi_{\bot\alpha}),      \eqno(14)
$$
$$
\langle \hat{b}_{\bot\alpha}\hat{b}_{\bot\alpha}\rangle=\int\frac{f_{\alpha}(\mathbf{k})}{(\xi_s^2k^2+\tau_{\bot}-\tau_{\|})^2}\frac{d^3k}{(2\pi)^3},         \eqno(15)
$$
where $\xi_{\bot}=\xi_s/\sqrt{\tau_{\bot}-\tau_{\|}}$.  In these formulae summation over the repeated index $\alpha$ is not assumed. The average product of the transverse fluctuations $\langle \hat{b}_{\bot\alpha}\hat{b}_{\bot\alpha}\rangle$ in a contrast to the longitudinal  $\langle \hat{b}_{\|}\hat{b}_{\|}\rangle$
practically does not depend on the temperature. This correction can be the most essential if the temperature is not too close to the transition, because the off-diagonal elements of the tensor   $\eta_{jl}$ are more sensitive to fluctuations of bending of strands. If $R_{\bot}$ is a length for a decay of correlations of  $\eta_{\alpha z}(\mathbf{r})$, then the argument analogous to that for the longitudinal components renders the following estimations:  $\langle \hat{b}_{\bot\alpha}\hat{b}_{\bot\alpha}\rangle\sim\frac{\mid\eta_{\alpha z}\mid^2}{(\tau_{\bot}-\tau_{\|})^2}$ at  $R_{\bot}\gg\xi_{\bot}$, and  $\langle \hat{b}_{\bot\alpha}\hat{b}_{\bot\alpha}\rangle\sim\frac{\mid\eta_{\alpha z}\mid^2R_{\bot}^3}{\xi_s^3(\sqrt{\tau_{\bot}-\tau_{\|}}}$ at  $R_{\bot}\ll\xi_{\bot}$.
The mixed averages $\langle \hat{b}_{\|}\hat{b}_{\bot\alpha}\rangle$ vanish because of the symmetry.

The average square of the amplitude $\Delta_P$ with accuracy up to the second order in $\eta_{jl}$ is found with the aid of projection Eq. (3) on $d_{\mu}m_j$.
Second order  correction to the transition temperature  $\tau_{zz}^{(1)}$ turns to zero the r.h.s. of the obtained equation. We don`t need this correction here.
Solution of the remaining equation renders:
$$
\langle\Delta_P^2\rangle=-\frac{\tau_{\|}}{\beta_{12345}}(1+\langle\hat{b}_l \hat{b}_l\rangle+2\langle \hat{b}_{\|} \hat{b}_{\|}\rangle)^{-1}\approx-\frac{\tau_{\|}}{\beta_{12345}}(1-\langle\hat{b}_l \hat{b}_l\rangle-2\langle \hat{b}_{\|} \hat{b}_{\|}\rangle),     \eqno(16)
$$
where $\hat{b}_l=b_l/\Delta_0$,  $\hat{b}_{\|}=b_{\|}/\Delta_0$.
The average order parameter is a good characteristic of a phase if corrections are small. According to the made above estimations  it means that the following conditions have to be met:  $\frac{\eta_{\alpha z}^2}{(\tau_{\bot}-\tau_{\|})^2}\ll 1$ at  $R_{\bot}\gg\xi_{\bot}$  or  $\frac{\mid\eta_{\alpha z}\mid^2R_{\bot}^3}{\xi_s^3(\sqrt{\tau_{\bot}-\tau_{\|}}}\ll 1$ at  $R_{\bot}\ll\xi_{\bot}$ for the transverse fluctuations  and  $\mid\tau_{\|}\mid\gg\left(\frac{R_{\|}^3}{\xi_s^3}\mid\eta_{zz}\mid^2\right)^2$ at $R_{\|}\ll\xi_{\|}$ or $\frac{\mid\eta_{zz}\mid^2}{\tau_{\|}^2}\ll 1$ at  $R_{\|}\gg\xi_{\|}$ for the longitudinal.

The quantities, entering these conditions can be roughly estimated with the aid of a model of long collinear cylinders randomly distributed in space with the two-dimensional density $n_2$ and specularly reflecting single-particle excitations  \cite{F18}. Within this model $(\tau_{\bot}-\tau_{\|})\sim n_2d\xi_s$, where  $d$ is the average diameter of the cylinders Using this relation we obtain $\frac{\mid\eta_{zz}\mid^2}{(\tau_{\bot}-\tau_{\|})^2}\sim\frac{1}{n_2\xi_s^2}\ll 1$. This is a natural condition of applicability of the mean-field approach. If the average distance between the strands is taken as $R$ the obtained condition can be rewritten as $\frac{R^2}{\xi_s^2}\ll 1$. According to the table in the ref.\cite{Dm-ms} this condition can be violated e.g. at pressures above $\sim$20 bar.
The condition for the limit $R_{\bot,\|}\ll\xi_{\bot,\|}$, i.e.  $\frac{\mid\eta_{\alpha z}\mid^2}{(\tau_{\bot}-\tau_{\|})^2}\frac{R_{\bot,\|}^3}{\xi_{\bot,\|}^3}\ll 1$ is less restrictive. Successful observation of the polar phase in nafen \cite{Dm-pp} shows that it has specific properties, which make it  a favorable material for investigation of this phase.

\section{NMR frequency shift}
For calculation of the NMR frequency shift we follow the argument of Ref.\cite{Dm-pp} but with account of  the fluctuations. The shift is due to the dipole energy \cite{VW}:
$$
U_D=\frac{3}{5}g_D(A_{jj}A_{\mu \mu}^\ast+A_{\mu j}A_{j \mu}^\ast),                                       \eqno(17)
$$
where $g_D$ is the dipole constant. Now $A_{\nu j}=\bar{A}_{\mu j}+a_{\mu j}$. According to the above argument the increment  $a_{\mu j}$ as well as the average order parameter  $\bar{A}_{\mu j}$ is proportional to the spin vector $d_{\mu}$. As a result the expression for the dipole energy can be represented as a product of the spin matrix $d_{j}d_{l}$ by the orbital matrix  $A_{j}A_{l}^\ast$. In the equations of spin dynamics the units can be chosen so that the gyromagnetic ratio for nuclei of  $^3$He - $g$ and magnetic susceptibility of $^3$He - $\chi$ were equal to each other. In these units the dipole energy has dimensionality of a square of frequency and it can be characterized by the square of the frequency of longitudinal oscillations $\Omega^2$. This frequrncy is different in different phases. Following Refs.  \cite{AG1,Dm-pp} we will normalize here all frequency shifts on the ratio $\Omega^2_A/\omega_L$, where $\Omega^2_A$ is a square of the longitudinal oscillation frequency in the bulk A-phase and $\omega_L$ - the Larmor frequency. In most experiments, including \cite{AG1,Dm-pp} these two frequencies satisfy a strong inequality  $\omega_L\gg\Omega_A$.    principal term of the expansion of the value of the relative NMR frequency shift $\Delta\omega/\omega_L$ over the small ratio  $\Omega^2/\omega_L^2$ is found by the averaging method of classical mechanics \cite{Mse,F76}. in application to the present problem it means that the spin matrix $d_j d_l$ has to be averaged over the fast (frequency $\sim\omega_L$) precession. Let us denote the result of  averaging as  $\overline{d_j d_l}\equiv D_{jl}$. Tensor  $D_{jl}$ depends on orientation of magnetic field and on the tipping angle $\beta$. The product of the  orbital parts has to be averaged over the ensemble of random tensors  $\eta_{jl}(\mathbf{r})$: $K_{jl}=\frac{1}{2}\langle A_jA_{l}^\ast+A_lA_{j}^\ast\rangle=\frac{1}{2}(\bar{A_j}\bar{A_{l}^\ast}+\bar{A_l}\bar{A_{j}}^\ast+\langle a_j a_l^\ast + a_l a_j^\ast\rangle)$. Tensor $K_{jl}$ has the following non-vanishing components:   $K_{xx}=\langle\Delta_P^2\rangle\langle\hat{b}_{\bot x}\hat{b}_{\bot x}\rangle$,  $K_{yy}=\langle\Delta_P^2\rangle\langle\hat{b}_{\bot y}\hat{b}_{\bot y}\rangle$   and      $K_{zz}=\langle\Delta_P^2\rangle(1+\langle\hat{b}_{\|}\hat{b}_{\|}\rangle)$. The resulting expression for the dipole energy has the form:
$$
\langle U_D\rangle=\frac{\Omega^2_A}{\Delta_A^2}D_{jl}K_{lj}.                                             \eqno(18)
$$
The amplitude $\Delta_A$ is borrowed from the definition of the order parameter of the A-phase ${A}_{\mu j}^{A}=\frac{\Delta_A}{\sqrt{2}}d_{\mu}(m_j+in_j)$.
Using the relation $2\omega_L\Delta\omega=-\frac{\partial\langle U_D\rangle}{\partial\cos\beta}$ and the values of components of $D_{jl}$ obtained in Ref. \cite{Dm-pp} we arrive at:
$$
\omega_L\Delta\omega=\Omega^2_A\frac{\beta_{245}}{\beta_{12345}}\times
$$
$$
\left\{\cos\beta-\frac{1}{4}\sin^2\mu
\left[2\langle\sin^2\Phi\rangle(1+\cos\beta)+5\cos\beta-1\right]\right\}(1-3\langle\hat{b}_{\bot x}\hat{b}_{\bot x}\rangle-2\langle\hat{b}_{\|}\hat{b}_{\|}\rangle).                                                    \eqno(19)
$$
Expression in the curly brackets coincides with that of Ref.\cite{Dm-pp}, it describes the dependence of the shift on a tipping angle $\beta$, an angle between the direction of magnetic field and average orientation of the strands  $\mu$,  and the relative phase $\Phi$ of rotation of the vector $d_j$ at a direction of spin $S$ with respect to the phase of precession of $S$. It means that the fluctuations do not change these dependencies. Only the overall coefficient is changed.

In particular, significant part of the results of the experiments \cite{Dm-pp} is represented in terms of the coefficient $K$ in the expression for the CV transverse  NMR shift  $2\omega_L\Delta\omega=K\Omega^2_A$ for the state of spin nematic ($\sin^2\Phi=0$) when the DC magnetic field is parallel to the strands  ($\sin^2\mu=0$):  . With an account of the contribution of fluctuations
$$
K=\frac{2\beta_{245}}{\beta_{12345}}(1-3\langle\hat{b}_{\bot x}\hat{b}_{\bot x}\rangle-2\langle\hat{b}_{\|}\hat{b}_{\|}\rangle).          \eqno(20)
$$
The ratio $\frac{\beta_{245}}{\beta_{12345}}$  enters this expression because for the uniform phases  $\frac{\Delta_P^2}{\Delta_A^2}=\frac{\beta_{245}}{\beta_{12345}}$.
Both correcting terms in Eq.(20) are negative. It means that the shift is always smaller than that in the uniform phase. The transverse fluctuations effect the value of the coefficient  $K$, the longitudinal introduce an additional temperature dependence, which is particularly important in the vicinity of the superfluid transition.

For a quantitative estimation of corrections to the value of $K$ a knowledge of specific  properties of the aerogels, e.g. of their structure factors is required. This limits possibility of the use of the value of $K$ for an unambiguous identification of superfluid phases. In particular, the value of $K$ can be a definitive indication of realization of the polar phase if $K$ is close to its maximum value, like it was in the experiment \cite{Dm-pp}. It means that the contribution of fluctuations to  $K$ is negligible. The converse argument does not work - the difference of $K$ from its maximum value does not mean that the observed phase is not polar \cite{AG1}.  It has to be remarked also that the obtained results can not be directly applied to interpretation of NMR spectra of states of the superfluid $^3$He with topological defects, in particular with the half-quantum vortices in the polar phase \cite{Autti}. For investigation of the effect of fluctuations on the satellite NMR lines, due to vortices a more involved analysis is needed. Such analysis  can be a subject of a separate work.
\section{Conclusions}
Nematic aerogels induce in the superfluid $^3$He a global orbital anisotropy. This anisotropy manifests itself in experiments on stabilization and investigation of the polar phase \cite{AG1,Dm-pp,Dm-ms,Autti}. One has to be aware that the effect of aerogel on the order parameter of superfluid $^3$He is spatially non-uniform. Except for the average global anisotropy a local random anisotropy is always present. The local anisotropy gives rise to spatial fluctuations of the order parameter, which contribute to macroscopic properties of a given phase of superfluid $^3$He, in particular on its NMR frequency shift. For the polar phase in magnetic field oriented parallel to the strands of aerogel the CV frequency shift is always smaller than the value expected from the mean-field calculations.
Relation between the effects of the global and the random local anisotropy is individual property of a particular aerogel. The available experimental data on structure of the practically used aerogels are not sufficient for making firm theoretical predictions. Further  experimental investigation of this question would be useful.

\section{Acknowledgements}
I thank V.V. Dmitriev for useful discussions and constructive criticism. I also thank the anonymous Referee for the interesting question.
The work was supported in part by the Program of the Presidium of RAS 1.4. "Actual problems of low temperature physics".


\begin{thebibliography}{99}

\bibitem{AG1} R. Sh. Askhadullin, V. V.Dmitriev, D. A. Krasnikhin, P. N. Martynov, A. A. Osipov, A.A. Senin, A.N.Yudin,   JETP Lett, \textbf{95}, 326 (2012).

\bibitem{Dm-pp}  V.V.Dmitriev, A.A. Senin, A.A. Soldatov , A.N.Yudin,  Phys. Rev. Lett.\textbf{115}, 165304 (2015)

\bibitem{Parp} N. Zhelev, M. Reichl, T. S. Abhilash, E. N. Smith, K. X. Nguen, E. J. Mueller, and J. M. Parpia, Nat. Commun. \textbf{7}, 12975 (2016).

\bibitem{AI} K. Aoyama and R. Ikeda, Phys. Rev. B \textbf{73}, 060504 (2006)

\bibitem{SF1} I. A. Fomin and E. V. Surovtsev, JETP Lett, \textbf{97}, 742 (2013).
.
\bibitem{fom3}  I. A. Fomin,  Zh.Exp.Teor.Fiz., \textbf{145}, 871 (2014).

\bibitem{nf} V. E. Asadchikov, R. Sh. Askhadullin, V. V. Volkov, V. V.Dmitriev, N. K. Kitaeva, P. N. Martynov, A. A. Osipov,
A. A. Senin, A. A. Soldatov, D. I. Chekrygina, and A. N.Yudin,  JETP Lett, \textbf{101}, 556 (2015).

\bibitem{VW} D.Vollhardt and P. Woelfle, The superfluid Phases of Helium 3, Taylor and Francis (1990).

\bibitem{LO} A. I. Larkin and Yu. N. Ovchinnikov, Zh.Exp.Teor.Fiz., {\bf 61}, 1221 (1971) [Sov. Phys. JETP,
{\bf 34}, 651 (1971)]

\bibitem{F18} I. A. Fomin, Zh.Exp.Teor.Fiz. \textbf{154}, 1034 (2018), (JETP, \textbf{127}, 933 (2018))

\bibitem{Mse} N. N. Moiseev, Asymptotic Methods of Nonlinear Mechanics  Ch.3, Nauka, Moscow (1969).

\bibitem{F76} I. A. Fomin, Zh.Exp.Teor.Fiz., \textbf{71}, 791 (1976).

\bibitem{Dm-ms}  V.V.Dmitriev, A.A. Soldatov , A.N.Yudin,  Phys. Rev. Lett.\textbf{120}, 075301 (2018)

\bibitem{Autti}  S. Autti,  V.V.Dmitriev, J. T. M$\ddot{a}$kinen, A. A. Soldatov, G. E. Volovik, A.N.Yudin, V. V. Zavjalov, and V. B. Eltsov, Phys. Rev. Lett.\textbf{117}, 255301 (2016)


\end{thebibliography}
\end{document}